\documentclass[11pt]{article}

\usepackage{amsmath,amssymb,amsthm}
\usepackage{geometry}
\usepackage{hyperref}
\usepackage{cite}
\usepackage{bbm}
\usepackage{mathtools}

\title{\large\bf Singularities, Entropy and the Arrow of Time, {\it or} Is CRT a Gauge Symmetry in Quantum Gravity?}

\author{T.~Banks\\[4pt]
\small NHETC and Department of Physics and Astronomy\\
\small Rutgers University, Piscataway, NJ 08854}

\date{}

\begin{document}
\maketitle

\begin{abstract}
We examine the question of whether the discrete transformation CRT is a gauge symmetry of ``Quantum Gravity".  Since the phrase in quotes is not yet completely well defined, we first try to define specific frameworks in which one might ask the question.  We find that the general answer is NO.  In asymptotically flat and AdS spaces, CRT is an asymptotic gauge symmetry in the same sense that the continuous parts of the Poincare/AdS isometry groups and scalar internal symmetries are. In eternal dS space it can be considered a spontaneously broken gauge symmetry if a certain ``Alice String" configuration is allowed in the Thermofield double Hilbert space.  This requires an extension of the conventional Chern-Simons rewriting of $2 + 1$ dimensional gravity. This interpretation only makes sense if we consider the underlying quantum mechanics to be time independent.  General quantum measurement and semi-classical gravitational restrictions on measuring devices put {\it a priori} limits on the existence of detectors with clocks that can actually measure proper time along a classical dS geodesic. 
\end{abstract}

\tableofcontents

\section{Introduction}
One of the most famous rigorous theorems about quantum field theories invariant under the component of the Poincare group connected to the identity, states that they are also invariant under the discrete Lorentz transformation CRT, which reflects time and one spatial coordinate axis, and changes the sign of every Lorentz invariant scalar charge.  This theorem can be extended to quantum field theories in curved space-times that admit natural deformation of the CRT transformation.  An immediate question that arises is whether some version of the theorem remains true when ``gravity is quantized", whatever that phrase is taken to mean.

For the author, and a large part of the high energy theory community, an accepted partial definition of the phrase "quantum gravity" would be models whose asymptotic expansion at weak coupling gives rise to one of the well defined superstring perturbation expansions in asymptotically flat space, and the AdS/CFT correspondence\cite{adscft}. Models in the former class are large $N$ limits of the BFSS matrix models\cite{bfsscomp} or models that can be obtained as large radius limits of AdS/CFT models.  In all of these cases CRT is definitely a symmetry of the theory, but it is {\it not} a gauge symmetry, since it acts on the gauge invariant Hilbert space.  This follows the familiar pattern in all gauge theories, that in systems with boundary, gauge transformations that act on the boundary are asymptotic gauge symmetries, which act on the boundary Hilbert space, because gauge transformations are defined to go to unity at the boundary.  In gravitational theories, at least so far, this paradigm only works for boundaries at infinite space-like distance and we call the resulting group of boundary diffeomorphisms {\it the asymptotic symmetry group}.  We've learned that internal symmetries of models of QG are also asymptotic gauge symmetries.   The same is true for CRT.  In AdS/CFT for example this can be seen by noting that we can do the usual Euclidean continuation of the boundary field theory on the Poincare patch, and CRT becomes part of the continuous Poincare group.  

\section{The Idealized Case of de Sitter Space}

Much of the literature on CRT as a gauge symmetry\cite{CRTgauge} is motivated by the idea of ``the path integral over closed universes"  .  What is usually meant by this is some kind of semi-classical expansion around de Sitter (dS) space.  As a consequence of general coordinate invariance, there are many different ways of thinking about what one means by the phrase semi-classical expansion.  Much of the current literature uses a loosely defined gravitational path integral, with little care taken about gauge fixing and whether the gauge choices involved make sense for the non-perturbative questions that are being asked.

Global classical dS space has an obvious CRT symmetry.  It is a time-like hyperboloid in a Minkowski space one dimension higher, and the RT symmetry of that Minkowski space, combined with reflection of all scalar symmetry charges, is a CRT symmetry.  It is relevant to note that, since the spatial sections of global dS are compact manifolds, any scalar charges that are integrals of densities coupled to bulk 1 form gauge fields automatically vanish.  If one believes that the rule from asymptotically flat and AdS space-times, that quantum gravity does not permit global 0-form symmetries, generalizes to closed universes, then the claim that the C part of CRT is a gauge symmetry follows.  

There is, however, a second way to interpret this same result, and indeed all of the conventional literature on QFT in dS space.  That is to view the Bunch-Davies\cite{BD} vacuum of dS space as the thermofield double (TFD) state of a quantum system living in one maximal static patch\cite{DKS}.  This mimicks Israel's\cite{israel} interpretation of the Kruskal extension of the Schwarzschild solution, which was generalized to AdS space by Maldacena\cite{malda}.  In this way of thinking about things, global dS space is just a trick for studying properties of the actual system in a single static patch.  The CRT transformation is just the usual relation between the two identical copies of the system with time reversed states, in the TFD.  One does {\it not} invoke an identification between the two static patches, but rather entangles them with fixed weight.

To make that last sentence more emphatic, we can refer to a paper of Parikh and Verlinde\cite{pv} who {\it did} consider orbifolding dS space by CRT, which is to say, treating CRT as a gauge transformation.  The resulting field theory Green's functions have singularities not only when two points coincide, but when a point approaches its image under CRT.   In interacting field theory, these singularities give rise to new divergences, which are not removed by the usual local counterterms\cite{tbwfmands}.  

Another possibility is that the TFD state represents a state with Higgsed CRT symmetry\cite{Susskind}. If that were the case, there would be a cosmic string in global dS space.  The holonomy around the string is the CRT transformation.  The cosmic string is Wilson line solution of $2 + 1$ dimensional dS gravity, living on the time-like geodesic between two static patches.  This would be an ``Alice $d - 3$ brane", a co-dimension two defect in space-time, the holonomy around which would be the CRT transformation. There is a problem with this idea if we write $2 + 1$ dimensional dS gravity as a Chern-Simons gauge theory. The action is {\it not} invariant under the extendions of $SO(3,1)$ to include the CRT transformation.  We would have to extend the gauge theory formulation to allow for defects that performed CRT and complex conjugated the gauge fields.   Susskind\cite{Susskind3} has proposed a kind of analog of a topological signature of Higgsing of CRT in $1 + 1$ dimensional de Sitter space.  It is not a cosmic string. 

The idea that CRT is spontaneously broken in cosmological solutions of gravity apparently originated in\cite{tbTCP} using the Wheeler-DeWitt formalism.  QFT in curved space-time follows from an expansion of the Wheeler-DeWitt wave function around a solution of the Hamilton-Jacobi constraint, which is quadratic in the canonical momentum to the instantaneous spatial metric.  There are two solutions, corresponding to the sign ambiguity in solving the quadratic constraint, one corresponding to an expanding, the other to a contracting cosmology.  CRT was interpreted as interchanging the two, and choosing to follow one branch of the wave function rather than the superposition was the analog of spontaneous breakdown of this symmetry.  

Exact eternal dS space is an equilibrium system, when treated in the ``approximation" of QFT in curved space-time.  Although the WD argument goes through for that geometry, the CRT symmetry discussed in\cite{tbTCP} is not the same transformation we have been discussing.  Furthermore, in the next section we'll introduce a quite different interpretation of cosmologies with a true Big Bang singularity and argue that the properly understood quantum theory has no CRT symmetry at all.  

In summary, to the extent that we view a model of external dS space as a time independent Hamiltonian quantum mechanics, with the global dS metric providing us with the TFD state corresponding the empty dS vacuum, then if we admit the existence of the Alice $d -3$ brane we can consider the CRT symmetry of the global dS manifold to be a spontaneously broken gauge symmetry.

\section{Singularities, Entropy and the Arrow of Time}

Let us quickly sketch the hydrodynamic view of quantum gravity, which follows from the work of\cite{ted95}\cite{carlip}\cite{solo}\cite{fsb}\cite{BZ}\cite{hst}\cite{hilbertbundles}. Each solution of the Lorentzian vacuum Einstein's equations defines the hydrodynamics of the state of some quantum system, called the {\it empty diamond state} of that geometry.   For us, ``Vacuum" Einstein solutions means that one factor space in the geometry is a maximally symmetric space-time.  Experience with string theory has taught (at least some of) us that different solutions correspond to different quantum mechanical systems.  Later in this section, we'll extend the discussion to include certain FRW cosmologies besides the maximally symmetric ones.  

The causal diamonds of the hydrodynamic geometry define the division of the quantum system up into subsystems with commuting finite dimensional operator algebras.  The work of the cited references tells us that the modular Hamiltonian of the density matrix in each diamond, in the empty diamond state, satisfies
\begin{equation} \langle K_{\diamond} \rangle = \langle (K_{\diamond} - \langle K_{\diamond} \rangle )^2 \rangle = \frac{A_{\diamond}}{4G_N} . \end{equation}  This relation holds without restriction for cosmological constant (c.c.) $\Lambda \geq 0$.  For $\Lambda < 0$ it holds only when the diamond is parametrically smaller than the AdS radius by a power of $(L_P / R_{AdS})$.  For $R_{\diamond} > R_{AdS_d}$, the fluctuations are smaller by a factor of $(d - 2)^{-1}$.  

These are the relations expected for a (cut-off) $1 + 1$ dimensional Conformal Field Theory (CFT), and using them we can build a causal time evolution operator inside nested causal diamonds, by mimicking half sided modular inclusion from algebraic quantum field theory (AQFT).   The future directed time evolution always has the expectation value of the modular Hamiltonian increasing.   In a CRT symmetric geometry we can always choose a past directed nesting, with diamond size increasing into the past, so there is no {\it a priori} choice for the direction of increase of entropy.  

One other general feature of this approach should be mentioned, which follows from assuming the Covariant Entropy Bound\cite{fsb}.  For positive c.c. and small diamonds for negative c.c., diamonds with localized excitations in them, {\it have less entropy than the empty diamond state}.  This is a very peculiar postulate for someone who is used to QFT, but it is the defining feature of QG.  One follows the trajectories of localized objects in the bulk by following patterns of frozen q-bits on the boundaries of nested diamonds.  The major virtue of this peculiar idea is that it fits the Schwarzschild dS entropy formula and the more general CEB argument alluded to above, and that it makes it easy to construct quantum mechanical models in which the scattering of localized objects leads to black hole production once the energy in a causal diamond exceeds the Schwarzschild threshhold  $E \sim R^{d-3} /G_N $.  

Given these hydrodynamic principles for interpreting gravity in terms of quantum mechanics we are now in a position to understand singularities and the arrow of time.  The Hawking-Penrose theorems tell us that singularities have to do with the vanishing of the area of causal diamonds in finite proper time along the geodesic in the diamond (and therefore along every time-like trajectory in the diamond).  We of course expect hydrodynamics to fail in general when entropy gets small\footnote{For systems with a unique ground state, quantized hydrodynamics and generalizations of it like Fermi liquid theory, can often describe the low lying micro-states, but in generic situations where hydrodynamics describes coarse grained flows, quantized hydrodynamics cannot describe micro-physics correctly.}.  Singularities refer to places where small diamonds are inevitable.

There is one situation, which is obviously non-singular: a quantum initial condition, which forces one to have small causal diamonds.  There is nothing singular about insisting that the initial conditions of a quantum system occurred a finite proper time in the past.  Assuming causality, and the CEB, this implies the breakdown of hydrodynamics/GR at the beginning of the universe.  Thus, in the hydrodynamic view of of the connection of GR to quantum mechanics, Cosmological Big Bang singularities are simply places where a description of the very earliest universe in terms of a set of non-interacting systems, each of which contains only a few q-bits, is appropriate, and the hydrodynamic description breaks down.  In\cite{tbwfmannelli} we showed how a large class of such small systems, with fairly random dynamics, could merge into the flat $p = \rho$ FRW universe as causal diamonds became large compared to the Planck scale. 

This connection between (fake) initial singularities and the thermodynamic arrow of time connects the cosmological and entropic time arrows, but does not yet lead to entropy of localized excitations.  The story of how those arise is more complex\cite{tbwfholocosm1}\cite{tbsacmb}\cite{tbwfsapbh}:  they are states of less than maximal entropy conditioned on the existence of a radiation dominated era and galaxies. 

Now let's consider inevitable final state singularities.  Here what we have to say is more controversial, but there are two situations where the story seems pretty clear.  The first are the singularities in black hole interiors.  The practical meaning of these singularities is that a detector inside the black hole can gather less and less quantum information about its environment the longer it delays the beginning of its experiment.  From the point of view of an outside detector, the infalling detector is equilibrating with the horizon degrees of freedom.  So an obvious interpretation of the infalling detector's experience is that the environment around it, and eventually its own degrees of freedom, are being equilibrated with the horizon.  This is not consistent with the locality of local field theory, but it is consistent with causality if quantum degrees of freedom live on causal diamond boundaries.  One fact that seems to validate this point of view is that the infall time to singularities is always roughly the same as the equilibration time measured by orbiting external observers.  

The other type of future space-like singularity that has been interpreted as equilibration of the degrees of freedom accessible to a local detector are the singularities in Coleman-deLuccia transitions.  There's a beautiful interpretation of transitions between two dS minima of an effective potential in terms of the principle of detailed balance\cite{tbdet}\cite{brwein}.  If one just shifts the same effective potential so that the higher minima has a small positive c.c. and the lower one has a negative c.c. which is larger in absolute value, then the Lorentzian evolution on the negative c.c. side has a future space-like singularity.  The maximal causal diamond in the crunching cosmology on the negative c.c. side has much smaller area than the static dS patch.  Again, the obvious interpretation is in terms of the principle of detailed balance.  The Lorentzian state on the negative c.c. side is a low entropy meta-stable state, into which the finite entropy dS system makes occasional excursions, but then rapidly returns to its much more probable equilibrium configuration.  

With these microscopic models of what underlies the hydrodynamic GR solutions, one sees that in the context of realistic cosmologies, CRT is simply not a symmetry of any kind.  It is intrinsically broken by the fact that we've imposed initial conditions on the quantum system that respect causality: small causal diamonds are not correlated.  Nonetheless the work of\cite{tbwfmannelli} shows that the most probable geometry associated with these principles is homogeneous, isotropic and flat in a coarse grained approximation.  The work of\cite{tbwfholocosm1}\cite{tbsacmb}\cite{tbwfsapbh} shows that plausible, lower entropy, initial conditions can account for the distributions of radiation and matter we actually see in our universe.  Singularities that are encountered in the future evolution of this cosmology are trapped inside black hole horizons and correspond to equilibration of transient local degrees of freedom with those horizons.  

\section{An Issue of Principle}

The notions of gauge symmetry, asymptotic gauge symmetry, and global symmetry, and the ``spontaneous breakdown" of each of these arose in well defined quantum field theories and lattice models, and have precise mathematical definitions.  Models of quantum gravity in asymptotically AdS space have similarly precise definitions, and models in asymptotically flat space have precise perturbative definitions, and in some cases precise definitions as limits of well defined quantum field theories.  From these definitions we've been able to prove that there are no global symmetries in QG and that symmetries that act on the Hilbert space can be viewed as asymptotic limits of local gauge transformations to all orders in semi-classical expansions\footnote{One cannot expect to go beyond such all orders statements since it's unclear there's a sharp definition of bulk geometry beyond the semi-classical expansion.}.  

If one accepts the conjecture\cite{tbwfds} that closed universe geometries with causal diamonds of finite maximal area correspond to quantum systems with finite dimensional Hilbert spaces, then it is impossible to make similar precise statements about symmetries for such systems.  A quantum measurement consists of entanglement between semi-classical q-bits of a {\it detector} and microscopic q-bits of the system being measured.  Semi-classical q-bits are two valued variables that have small fluctuations, inversely proportional to some power of a large integer $N$, and obey the sum over histories rule for probabilities (rather than amplitudes) with exponential accuracy in $N$.  A mathematical construction of such q-bits in cut-off QFT is easy: we simply average local variables over large regions whose size in microscopic units scales like $N$.  However, semi-classical gravity shows us that only a sub-leading power of the entropy in a finite area causal diamond is well described by bulk quantum field theory. Most of it is contained in degrees of freedom that behave like the horizon of a black hole. For non-negative cosmological constant or scales smaller than the AdS radius for negative c.c., they scramble quantum information rapidly and support only a tiny number of semi-classical q-bits.  {\it So no mathematical theory of a universe with only finite area causal diamonds can ever be tested experimentally with any sort of precision, by measurements inside that universe.}  The implication of this is that we can modify any theory of dS space that someone builds, in a way that agrees with gross semi-classical predictions, without fear of contradiction until an actual experiment falsifies the modification.

Consider for example models that attribute a time independent Hamiltonian to dS space.  Whatever the Hamiltonian is, we can modify it by allowing its parameters to vary slowly with time in a random manner.  By making the variation slow enough, we can evade any experimental contradiction.  We can also add small terms to the Hamiltonian that don't modify the semi-classical physics or any {\it current} experimental predictions.  The {\it a priori} arguments above show that no matter how many years we wait and how precise our experimental techniques become, we will still be able to make indistinguishable modifications of the theory.  It should be obvious that these in principle unmeasurable terms can violate any symmetry we wish, or any preconception about whether it is gauged or spontaneously broken.

Explicit illustrations of these remarks can be found in the low dimensional models of\cite{satbwfds3}\cite{satbjtds}, where we exhibited models of dS gravity in $2 + 1$ and $ 1 + 1$ dimensions, which matched many semi-classical properties of solutions of the Einstein field equations coupled to certain kinds of matter.  The models had large numbers of free parameters, none of which could be determined from the semi-classical physics.  The actual quantum physics was chaotic and lived on the boundaries of causal diamonds.  It was not amenable to detection by bulk detectors.  Something similar happened in the model of\cite{tbpdbz} which purports to describe the propagation of massless fermions in the long throat of an extremal Reissner-Nordstrom black hole.  The fermions in the throat are transient excitations, which eventually fall into a horizon described by a model picked from the SYK ensemble.  It's plausible to speculate that the random coupling in this ensemble might arise from the many different ways of making such a black hole in four dimension string theories.  In principle, some particular string model might be able to measure a particular set of SYK couplings.  The finite lifetime, finite entropy fermion system in the throat cannot do so.  

There is one possible mathematical route to a more precise definition of CRT in dS space.   We can view a sequence of models of dS space with discrete c.c. tending to zero as approximations to some particular model of QG in asymptotically flat space.  It's then clear that the CRT symmetry {\it of a single horizon patch of dS space} formally converges to the CRT symmetry of Minkowski space and should be considered an unbroken asymptotic gauge symmetry.  This is not the same as the CRT transformation that we considered above as a possible spontaneously broken gauge symmetry of a theory of eternal dS space.  

\section{Conclusions}

We've briefly reviewed various ideas about the role of CRT in models of quantum gravity.
In well defined models with asymptotically infinite boundaries it is an asymptotic gauge symmetry, with non-trivial action on states in the Hilbert space.   We also argued, using the hydrodynamic interpretation of the relationship between GR and quantum mechanics, that in a realistic Big Bang cosmology, CRT was simply not a symmetry at all.  Initial singularities are associated with a finite past where hydrodynamic approximations break down.  Final singularities are all associated with the breakdown of hydrodynamic approximations for particular local detector systems, which get trapped in low entropy subsystems.  The illusion that they are propagating in a hydrodynamic space-time does not survive the equilibration of the q-bits that make up that space-time with a much larger system.  We conjectured that the resolution of {\it all} future space-like singularities has this kind of an explanation.

For the special case of eternal dS space, or ``the gravitational path integral over closed universes" to which it is supposed to be the semi-classical approximation, we advocated for the interpretation\cite{DKS} of the global dS geometry as a description of the TFD state of a system that lives in a single static patch.  The CRT transformation does not act on the actual quantum system representing dS space, but is the usual ``swap" transformation of a general TFD state.  Although if we allow the Alice $d - 3$ brane configuration as a state in the TFD Hilbert space, there's a sense in which it resembles a spontaneously broken gauge symmetry, this is only true to the extent that the static Hamiltonian is actually a description of time evolution.  We argued that if dS space really has a finite number of quantum states, there is no way in principle that a detector in that space-time could register the existence of either the static Killing vector or the CRT transformation.

If we view a sequence of dS models with c.c. tending to zero as approximations to a model in asymptotically flat space, then the CRT transformation of a single static patch becomes the CRT transformation of the limiting flat space and converges to an asymptotic gauge symmetry in the limit of vanishing c.c..  
\section*{Acknowledgments}

The author thanks Claude, Gemini, L. Susskind, and D. Harlow for useful discussions of the material in this paper.  

\end{document}